\documentclass[draftcls, onecolumn, letterpaper, 11pt]{IEEEtran}

\usepackage[dvips]{graphicx}
\graphicspath{{../eps/}}
\DeclareGraphicsExtensions{.eps}

\usepackage{amsopn}
\usepackage{multirow}
\usepackage{amsmath}
\usepackage{amsfonts}
\usepackage{amssymb}
\usepackage{longtable}

\usepackage{ifthen}
\newboolean{ONECOLUMN}
\setboolean{ONECOLUMN}{true}
\usepackage{verbatim}\begin{comment}For comment spanned over multiple lines.\end{comment}
\ifCLASSOPTIONcompsoc\usepackage[tight,normalsize,sf,SF]{subfigure}\else\usepackage[tight,footnotesize]{subfigure}\fi

\usepackage[ruled, vlined]{algorithm2e}
\usepackage{hyperref}

\begin{document}
\title{2D Sparse Signal Recovery via 2D Orthogonal Matching Pursuit}
\author{Yong~Fang, Bormin Huang, and Jiaji Wu% <-this % stops a space
\thanks{This research was supported by National Science Foundation of China (NSFC) (grant no. 61001100) and Provincial Science Foundation of Shaanxi, China (grant no. 2010K06-15).}% <-this % stops a space
\thanks{Y. Fang is with College of Information Engineering, Northwest A\&F University, Shaanxi Yangling 712100, China (email: yfang79@gmail.com).}
\thanks{B. Huang is with Space Science and Engineering Center, University of Wisconsin-Madison, 1225 W. Dayton St. Madison, WI 53706, USA (email: bormin@ssec.wisc.edu).}
\thanks{J. Wu is with Key Laboratory of Intelligent Perception and Image Understanding of Ministry of Education of China, Xidian University, Xi'an 710071, China (email: wujj@mail.xidian.edu.cn).}}
\maketitle% make the title area

\begin{abstract}
	Recovery algorithms play a key role in compressive sampling (CS). Most of current CS recovery algorithms are originally designed for one-dimensional (1D) signal, while many practical signals are two-dimensional (2D). By utilizing 2D separable sampling, 2D signal recovery problem can be converted into 1D signal recovery problem so that ordinary 1D recovery algorithms, e.g. orthogonal matching pursuit (OMP), can be applied directly. However, even with 2D separable sampling, the memory usage and complexity at the decoder is still high. This paper develops a novel recovery algorithm called 2D-OMP, which is an extension of 1D-OMP. In the 2D-OMP, each atom in the dictionary is a matrix. At each iteration, the decoder projects the sample matrix onto 2D atoms to select the best matched atom, and then renews the weights for all the already selected atoms via the least squares. We show that 2D-OMP is in fact equivalent to 1D-OMP, but it reduces recovery complexity and memory usage significantly. What's more important, by utilizing the same methodology used in this paper, one can even obtain higher dimensional OMP (say 3D-OMP, etc.) with ease.
\end{abstract}


%%%%%%%%%%%%%%%%%%%%%%%%%%%%%%%%%%%%%%%%%%%%%%%%%%%%%%%%%%%%%%%%%%%%%%%%%%%%%%%%%%%%%%%%%%%%%%%%%%%%%%%%%%%%%%%%%%%%%%%%%%%%%%%%%%%%%%%%%%%%%%%%%%%%%%%%%%%%%%%%%%%%%%%%%
\section{Introduction}\label{sec:intro}
	\IEEEPARstart{L}{et} $\boldsymbol{x} \in \mathbb{R}^n$ be a one-dimensional (1D) signal and $\mathbf{\Psi} \in \mathbb{R}^{n \times n}$ an orthonormal transform matrix, where $\mathbb{R}$ is the set of real numbers. If $\boldsymbol{x} = \mathbf{\Psi}\boldsymbol{z}$ and there are only $k \ll n$ spikes (nonzero entries) in $\boldsymbol{z}$, we say that $\boldsymbol{x}$ is $k$-sparse in $\mathbf{\Psi}$ domain. We sample $\boldsymbol{x}$ by $\mathbf{\Phi} \in \mathbb{R}^{m \times n}$ ($k < m < n$) to get $\boldsymbol{y} = \mathbf{\Phi}\boldsymbol{x} = \mathbf{A}\boldsymbol{z} \in \mathbb{R}^{m}$, where $\mathbf{A} = \mathbf{\Phi}\mathbf{\Psi}$. If $\mathbf{\Phi}$ obeys the order-$k$ restricted isometry property (RIP) and has low coherence with $\mathbf{\Psi}$, then $\boldsymbol{z}$ (and in turn $\boldsymbol{x}$) can be effectively recovered from $\boldsymbol{y}$ \cite{Candes_IEEE_TIT_06_02, Candes_IEEE_TIT_06_12, Donoho_IEEE_TIT_06_04}. Many algorithms have been proposed to recover $\boldsymbol{x}$ from its random sample $\boldsymbol{y}$, e.g. linear programming (LP) \cite{Candes_IEEE_TIT_05_12} and orthogonal matching pursuit (OMP) \cite{Tropp_IEEE_TIT_07_12}. For a detailed overview on recovery algorithms, please refer to \cite{Pope_Master_Thesis_09_02}.

	In practice, many signals, e.g. image, video, etc, are two-dimensional (2D). A straightforward implementation of 2D compressive sampling (CS) is to stretch 2D matrices into 1D vectors. However, such direct stretching increases exponentially the complexity and memory usage at both encoder and decoder. An alternative to 1D stretching is to sample rows and columns of 2D signals independently by using separable operators \cite{Rivenson_IEEE_SPL_09}. Through 2D separable sampling, encoding complexity is exponentially reduced. However, as the recovery problem is converted into a standard 1D $\ell_1$-minimization problem, decoding complexity is still very high.

	As a representative sparse signal recovery algorithm, the OMP achieves good performance with low complexity. The OMP is originally designed for 1D signal recovery. To reduce the complexity of 2D signal recovery, this paper extends the 1D-OMP to obtain the 2D-OMP. We prove that with 2D separable sampling, 2D-OMP is in fact equivalent to 1D-OMP, so that both algorithms will output exactly the same results. However, the complexity and memory usage of 2D-OMP is much lower than that of 1D-OMP. Thus, 2D-OMP can be used as an alternative to 1D-OMP in 2D sparse signal recovery.

	This paper is arranged as follows. Section \ref{sec:1D_OMP} first briefly reviews the principles of 2D separable sampling and 1D-OMP, and then makes a detailed analysis on the complexity of 1D-OMP. In Section \ref{sec:2D_OMP}, we deduce the 2D-OMP algorithm, reveal the equivalence of 2D-OMP to 1D-OMP, and compare the complexity and memory usage of 2D-OMP with that of 1D-OMP. In Section \ref{sec:results}, simulation results are reported. Finally, Section \ref{sec:conclusion} concludes this paper.

%%%%%%%%%%%%%%%%%%%%%%%%%%%%%%%%%%%%%%%%%%%%%%%%%%%%%%%%%%%%%%%%%%%%%%%%%%%%%%%%%
%%%%%%%%%%%%%%%%%%%%%%%%%%%%%%%%%%%%%%%%%%%%%%%%%%%%%%%%%%%%%%%%%%%%%%%%%%%%%%%%%
%%%%%%%%%%%%%%%%%%%%%%%%%%%%%%%%%%%%%%%%%%%%%%%%%%%%%%%%%%%%%%%%%%%%%%%%%%%%%%%%%
\section{1D Orthogonal Matching Pursuit with 2D Separable Sampling}\label{sec:1D_OMP}

\subsection{2D Separable Sampling}
	The principle of 2D separable sampling \cite{Rivenson_IEEE_SPL_09} is as follows. Let $\mathbf{X} \in \mathbb{R}^{n \times n}$ be a 2D signal which is $k$-sparse in $\mathbf{\Psi}$ domain, i.e. $\mathbf{X} = \mathbf{\Psi} \mathbf{Z} \mathbf{\Psi}^\mathrm{T}$ and there are only $k \ll n^2$ spikes in $\mathbf{Z}$, where $(\cdot)^{\mathrm{T}}$ denotes the transpose. For simplicity, we use the same operator $\mathbf{\Phi}$ to sample the rows and columns of $\mathbf{X}$ independently to get $\mathbf{Y} = \mathbf{\Phi} \mathbf{X} \mathbf{\Phi}^{\mathrm{T}} = \mathbf{A} \mathbf{Z} \mathbf{A}^{\mathrm{T}} \in \mathbb{R}^{m \times m}$. Let $\boldsymbol{y} \in \mathbb{R}^{m^2}$ be the 1D stretched vector of $\mathbf{Y}$ and  $\boldsymbol{z} \in \mathbb{R}^{n^2}$ the 1D stretched vector of $\mathbf{Z}$. It was proved that
\begin{equation}	
	\boldsymbol{y} = (\mathbf{\Phi}\otimes\mathbf{\Phi}) (\mathbf{\Psi}\otimes\mathbf{\Psi}) \boldsymbol{z} = \mathbf{\Omega}\boldsymbol{z},
\end{equation}
where $\otimes$ denotes the Kronecker product \cite{Rivenson_IEEE_SPL_09}. It is easy to prove $(\mathbf{\Phi}\otimes\mathbf{\Phi}) (\mathbf{\Psi}\otimes\mathbf{\Psi}) = (\mathbf{\Phi}\mathbf{\Psi}) \otimes (\mathbf{\Phi}\mathbf{\Psi})$, hence $\mathbf{\Omega} = \mathbf{A}\otimes\mathbf{A}$. Now this is just a standard 1D sparse signal recovery problem which can be attacked by the OMP.

\subsection{1D Orthogonal Matching Pursuit}\label{sec:1d_algorithm}

\begin{algorithm}[t]
\label{alg:1D_OMP}
\caption{1D Orthogonal Matching Pursuit with 2D Separable Sampling}
\textbf{Input}:\
\begin{itemize}
	\item $\mathbf{\Omega} \in \mathbb{R}^{m^2 \times n^2}$: sampling matrix
 	\item $\boldsymbol{y} \in \mathbb{R}^{m^2}$: sample 
 	\item $k$: sparsity level
\end{itemize}

\textbf{Output}:\
\begin{itemize}
	\item $\tilde{\boldsymbol{z}} \in \mathbb{R}^{n^2}$: reconstruction of the ideal signal $\boldsymbol{z}$
\end{itemize}

\textbf{Auxiliary Variables}:\ 
\begin{itemize}
	\item $\boldsymbol{r} \in \mathbb{R}^{m^2}$: residual
	\item $\boldsymbol{i}$: set of the indices of atoms that are allowed to be selected in the future 
\end{itemize}

\textbf{Initialization}:\
\begin{itemize}
	\item $\boldsymbol{r} \gets \boldsymbol{y}$
	\item $\boldsymbol{i} \gets \{1, 2, \cdots, n^2\}$
\end{itemize}

\For{$t\leftarrow 1$ \KwTo $k$}{
	$i_t \gets \arg\underset{{i'}\in\boldsymbol{i}}{\max} \frac{\left|\left\langle\boldsymbol{r}, \boldsymbol{\omega}_{i'}\right\rangle\right|}
	 												 	 {\left\|\boldsymbol{\omega}_{i'}\right\|_2}$\;
	$\boldsymbol{i} \gets \boldsymbol{i} \setminus i_t$\;	 												 	 
	$\hat{\boldsymbol{u}} \gets \arg\underset{\boldsymbol{u}}{\min}\left\|\boldsymbol{y} -	\sum_{t'=1}^{t}{u_{t'}\boldsymbol{\omega}_{i_{t'}}}\right\|_2$\;
	$\boldsymbol{r} \gets \boldsymbol{y} - \sum_{t'=1}^{t}{\hat{u}_{t'}\boldsymbol{\omega}_{i_{t'}}}$\;
}

\For{$t\leftarrow 1$ \KwTo $k$}{
	$\tilde{z}_{i_t} \gets \hat{u}_t$\;
}
\end{algorithm}

	Let $\mathbf{\Omega} = (\boldsymbol{\omega}_1, \cdots, \boldsymbol{\omega}_{n^2})$, where $\boldsymbol{\omega}_{i}\in\mathbb{R}^{m^2}$ is the $i$-th column of $\mathbf{\Omega}$. We call $\mathbf{\Omega}$ the \textit{dictionary} and $\boldsymbol{\omega}_i$ an \textit{atom}. The main idea of 1D-OMP is to represent $\boldsymbol{y}$ as a weighted sum of as few atoms as possible.
	
	Algorithm \ref{alg:1D_OMP} gives main steps of 1D-OMP. To implement 1D-OMP, we need two auxiliary variables. First, to avoid atom reselection, set $\boldsymbol{i}$ is defined to record the indices of those atoms that are allowed to be selected in the future (excluding those already selected atoms). Second, vector $\boldsymbol{r} \in \mathbb{R}^{m^2}$ is defined to hold the residual after removing the selected atoms from $\boldsymbol{y}$. Initially, $\boldsymbol{r}$ is set to $\boldsymbol{y}$. Then at each iteration, the decoder picks from the dictionary the atom that best matches the residual and then renews the weights for all the already selected atoms via the least squares.
	
\subsection{Complexity Analysis}\label{sec:1d_analysis}
	We decompose the iteration of 1D-OMP into the following steps and analyze its complexity step by step.
	
\subsubsection{Project} The projection of residual $\boldsymbol{r}$ onto atom $\boldsymbol{\omega}_{i'}$ is $\left\langle \boldsymbol{r}, \boldsymbol{\omega}_{i'} \right\rangle / \left\| \boldsymbol{\omega}_{i'} \right\|_2$, where $\left\langle \cdot, \cdot \right\rangle$ denotes the inner product between two vectors and $\left\| \cdot \right\|_2$ denotes the $\ell_2$-norm of a vector. Let $\boldsymbol{\rho} = (\left\|\boldsymbol{\omega}_1\right\|_2, \cdots, \left\|\boldsymbol{\omega}_{n^2}\right\|_2)^\mathrm{T}$, then this step can be implemented by $\mathbf{\Omega}^\mathrm{T}\boldsymbol{r} ./ \boldsymbol{\rho}$, where $./$ denotes dot division. The complexity of this step is dominated by $(n^2 \times m^2) \times (m^2 \times 1)$ matrix-vector multiplication. Hence the complexity of this step is $O(m^2 n^2)$. 

\subsubsection{Select Best Matched Unselected Atom} This step selects from unselected atoms the atom with the maximal absolute value of projection. As there are $n^2$ atoms and $k \ll n^2$, the complexity of this step is approximately $O(n^2)$, negligible compared with the $\textit{Project}$ step.

\subsubsection{Renew Weights} Let $\mathbf{\Omega}_{\boldsymbol{i}} = (\boldsymbol{\omega}_{i_1}, \cdots, \boldsymbol{\omega}_{i_t})$, then $\sum_{t'=1}^{t}{u_{t'}\boldsymbol{\omega}_{i_{t'}}} = \mathbf{\Omega}_{\boldsymbol{i}}\boldsymbol{u}$. According to linear algebra,
\begin{equation}
	\arg\underset{\boldsymbol{u}}{\min} \left\| \boldsymbol{y} - \mathbf{\Omega}_{\boldsymbol{i}}\boldsymbol{u} \right\|_2 = \mathbf{Q}^{-1} \boldsymbol{g},
\end{equation}
where $\mathbf{Q} = \mathbf{\Omega}_{\boldsymbol{i}}^\mathrm{T}\mathbf{\Omega}_{\boldsymbol{i}} \in \mathbb{R}^{t \times t}$ and $\boldsymbol{g} = \mathbf{\Omega}_{\boldsymbol{i}}^\mathrm{T} \boldsymbol{y} \in \mathbb{R}^{t}$. The complexity to calculate $\mathbf{Q}$ and $\boldsymbol{g}$ depends on $t$. For $t \leq k \ll n^2$, the complexity of this step is negligible compared with the $\textit{Project}$ step. 

\subsubsection{Update Residual} The complexity of this step depends on $t$. For $t \leq k \ll n^2$, the complexity of this step is negligible compared with the $\textit{Project}$ step.

Based on the above analysis, we conclude that the complexity of 1D-OMP is dominated by the $\textit{Project}$ step and its complexity is $O(m^2 n^2)$.

%%%%%%%%%%%%%%%%%%%%%%%%%%%%%%%%%%%%%%%%%%%%%%%%%%%%%%%%%%%%%%%%%%%%%%%%%%%%%%%%%
%%%%%%%%%%%%%%%%%%%%%%%%%%%%%%%%%%%%%%%%%%%%%%%%%%%%%%%%%%%%%%%%%%%%%%%%%%%%%%%%%
%%%%%%%%%%%%%%%%%%%%%%%%%%%%%%%%%%%%%%%%%%%%%%%%%%%%%%%%%%%%%%%%%%%%%%%%%%%%%%%%%
\section{2D Orthogonal Matching Pursuit}\label{sec:2D_OMP}
	This section develops the 2D-OMP algorithm whose main idea is to represent 2D signal $\mathbf{Y}$ as a weighted sum of 2D atoms that are selected from an over-complete dictionary. We first redefine the concepts of atom, dictionary, and projection for 2D signals. Then we give the 2D-OMP algorithm. We reveal the equivalence of 2D-OMP to 1D-OMP and compare the complexity and memory usage of 2D-OMP with that of 1D-OMP.

\subsection{Definition}
	Let $\mathbf{X} \in \mathbb{R}^{n \times n}$ be a 2D signal that is $k$-sparse in $\mathbf{\Psi}$ domain and $\mathbf{Y} = \mathbf{\Phi} \mathbf{X} \mathbf{\Phi}^{\mathrm{T}} = \mathbf{A} \mathbf{Z} \mathbf{A}^{\mathrm{T}} \in \mathbb{R}^{m \times m}$. Let $\mathbf{A} = (\boldsymbol{a}_{1}, \cdots, \boldsymbol{a}_{n})$, where $\boldsymbol{a}_{i}$ is the $i$-th column of $\mathbf{A}$. We redefine dictionary, atom, and projection as follows.
	
\subsubsection{Dictionary and Atom}
	In the 2D-OMP, the dictionary contains $n^2$ atoms and each atom is an $m \times m$ matrix. Let $\mathbf{B}_{i,j}$ be the ($i$, $j$)-th atom, then $\mathbf{B}_{i,j}$ is the outer product of $\boldsymbol{a}_{i}$ and $\boldsymbol{a}_{j}$
\begin{equation}
	\mathbf{B}_{i,j} = \boldsymbol{a}_{i} \boldsymbol{a}_{j}^\mathrm{T} =
	\left(
		\begin{array}{ccc}
			a_{1i}a_{1j} 	& \cdots 	& a_{1i}a_{mj}\\
			\vdots 			& \vdots 	& \vdots\\
			a_{mi}a_{1j} 	& \cdots 	& a_{mi}a_{mj}
		\end{array}
	\right).
\label{eq:2D_atom}	
\end{equation}
Now $\mathbf{Y}$ can be represented by the weighted sum of $\mathbf{B}_{i,j}$, i.e.
\begin{equation}
	\mathbf{Y} = \sum_{i=1}^{n}{\sum_{j=1}^{n}{z_{i,j}\mathbf{B}_{i,j}}}.
\label{eq:2d_project}	
\end{equation}

\subsubsection{Projection}	
	The projection of $\mathbf{Y}$ onto $\mathbf{B}_{i,j}$ is
\begin{equation}
	\frac{\left\langle\mathbf{Y}, \mathbf{B}_{i,j}\right\rangle}{\left\|\mathbf{B}_{i,j}\right\|_2},
\end{equation}
where $\left\langle\mathbf{Y}, \mathbf{B}_{i,j}\right\rangle \triangleq \boldsymbol{a}_i^\mathrm{T} \mathbf{Y} \boldsymbol{a}_j$ and $\left\|\mathbf{B}_{i,j}\right\|_2$ is the Frobenius norm of $\mathbf{B}_{i,j}$, i.e.
\begin{equation}
	\left\|\mathbf{B}_{i,j}\right\|_2 \triangleq \sqrt{\sum_{i'=1}^{m}{\sum_{j'=1}^{m}{(a_{i'i} a_{j'j})^2}}} = \left\|\boldsymbol{a}_i\right\|_2 \left\|\boldsymbol{a}_j\right\|_2.
\end{equation}

\subsection{Algorithm Description}
\begin{algorithm}[t]
\label{alg:2D_OMP}
\caption{2D Orthogonal Matching Pursuit}
\textbf{Input}:\
\begin{itemize}
	\item $\mathbf{A} \in \mathbb{R}^{m \times n}$: sampling matrix
 	\item $\mathbf{Y} \in \mathbb{R}^{m \times m}$: sample 
 	\item $k$: sparsity level
\end{itemize}

\textbf{Output}:\
\begin{itemize}
	\item $\tilde{\mathbf{Z}} \in \mathbb{R}^{n \times n}$: reconstruction of the ideal signal $\mathbf{Z}$
\end{itemize}

\textbf{Variable}:\
\begin{itemize}
	\item $\mathbf{R} \in \mathbb{R}^{m \times m}$: residual
	\item $(\boldsymbol{i}$, $\boldsymbol{j})$: set of the coordinates of atoms that are allowed to be selected in the future, $\boldsymbol{i}$ for row indices and $\boldsymbol{j}$ for column indices
\end{itemize}

\textbf{Initialization}:\
\begin{itemize}
	\item $\mathbf{R} \gets \mathbf{Y}$
	\item $(\boldsymbol{i}$, $\boldsymbol{j}) \gets \{(1,1), (1,2), \cdots, (n, n)\}$
\end{itemize}

\For{$t\leftarrow 1$ \KwTo $k$}{
	$(i_t, j_t) \gets \arg\underset{(i',j')\in(\boldsymbol{i}, \boldsymbol{j})}{\max}\frac{\left|\left\langle\mathbf{R}, \mathbf{B}_{i',j'}\right\rangle\right|}
																{\left\|\mathbf{B}_{i',j'}\right\|_2}$\;
	$(\boldsymbol{i}, \boldsymbol{j}) \gets (\boldsymbol{i}, \boldsymbol{j}) \setminus (i_t, j_t)$\;
	$\hat{\boldsymbol{u}} \gets \arg\underset{\boldsymbol{u}}{\min}	\left\|\mathbf{Y} - \sum_{t'=1}^{t}{u_{t'}\mathbf{B}_{i_{t'},j_{t'}}} \right\|_2$\;
	$\mathbf{R} \gets \mathbf{Y} - \sum_{t'=1}^{t}{\hat{u}_{t'}\mathbf{B}_{i_{t'},j_{t'}}}$\;
}

\For{$t\leftarrow 1$ \KwTo $k$}{
	$\tilde{z}_{i_t,j_t} \gets \hat{u}_t$\;
}
\end{algorithm}

	Algorithm \ref{alg:2D_OMP} gives main steps of 2D-OMP. To implement the 2D-OMP algorithm, we also need two auxiliary variables. First, to avoid atom reselection, set $(\boldsymbol{i}, \boldsymbol{j})$ is defined to record the coordinates of those atoms that are allowed to be selected in the future (excluding those already selected atoms), where $\boldsymbol{i}$ for row indices and $\boldsymbol{j}$ for column indices. Second, $\mathbf{R} \in \mathbb{R}^{m \times m}$ is defined to hold the residual after removing the selected atoms from $\mathbf{Y}$. Initially, $\mathbf{R}$ is set to $\mathbf{Y}$. Then at each iteration, the decoder first searches for the best matched atom in the dictionary and then renews the weights for all the already selected atoms via the least squares.
	
\subsubsection{Least Squares}
	The weighted sum of $t$ selected atoms constructs an approximation to $\mathbf{Y}$. Let
\begin{equation}	
	\mathbf{R} = \mathbf{Y} - \sum_{t'=1}^{t}{u_{i_{t'},j_{t'}}\mathbf{B}_{i_{t'},j_{t'}}}.
	\label{eq:R}
\end{equation}	
We model the problem as finding the optimal $\boldsymbol{u} = (u_{i_1,j_1}, \cdots, u_{i_t,j_t})^\mathrm{T}$ that minimizes the Frobenius norm of $\mathbf{R}$, which is in fact equivalent to the least squares problem. As $\left\| \mathbf{R} \right\|_2^2 = \mathrm{tr}(\mathbf{R}\mathbf{R}^\mathrm{T})$, where $\mathrm{tr}(\cdot)$ is the trace of a matrix, the problem is equivalent to 
\begin{equation}	
	\hat{\boldsymbol{u}} = \arg \underset{\boldsymbol{u}}{\min}\, \mathrm{tr}(\mathbf{R}\mathbf{R}^\mathrm{T}). 
\end{equation}	
Using (\ref{eq:R}) and $\mathrm{tr}((\cdot)^{\mathrm{T}}) = \mathrm{tr}(\cdot)$, we have
\begin{eqnarray}
\mathrm{tr}(\mathbf{R}\mathbf{R}^\mathrm{T}) 
	&=& \mathrm{tr}(\mathbf{Y}\mathbf{Y}^\mathrm{T}) - 
		2\sum_{t'=1}^{t}{u_{i_{t'},j_{t'}} \mathrm{tr}(\mathbf{Y}\mathbf{B}_{i_{t'},j_{t'}}^\mathrm{T})} + \nonumber\\
	& &	\sum_{t'=1}^{t}{\sum_{s'=1}^{t}{u_{i_{t'},j_{t'}}u_{i_{s'},j_{s'}}\mathrm{tr}(\mathbf{B}_{i_{t'},j_{t'}}\mathbf{B}_{i_{s'},j_{s'}}^\mathrm{T})}}\nonumber\\
	&=& \left\| \mathbf{Y} \right\|_2^2 + \boldsymbol{u}^\mathrm{T} \mathbf{H} \boldsymbol{u} - 
		2\boldsymbol{f}^\mathrm{T} \boldsymbol{u},
\end{eqnarray}
where 
\begin{equation} 
\mathbf{H} = \left(
\begin{array}{ccc}
\mathrm{tr}(\mathbf{B}_{i_1,j_1}\mathbf{B}_{i_1,j_1}^\mathrm{T}) & \cdots & \mathrm{tr}(\mathbf{B}_{i_1,j_1}\mathbf{B}_{i_t,j_t}^\mathrm{T})\\
\vdots & \vdots & \vdots\\
\mathrm{tr}(\mathbf{B}_{i_t,j_t}\mathbf{B}_{i_1,j_1}^\mathrm{T}) & \cdots & \mathrm{tr}(\mathbf{B}_{i_t,j_t}\mathbf{B}_{i_t,j_t}^\mathrm{T})
\end{array}
\right)
\end{equation}
and
\begin{equation} 
	\boldsymbol{f} = (\mathrm{tr}(\mathbf{Y} \mathbf{B}_{i_1,j_1}^\mathrm{T}), \cdots, \mathrm{tr}(\mathbf{Y} \mathbf{B}_{i_t,j_t}^\mathrm{T}))^\mathrm{T}.
\end{equation} 
When $\mathrm{tr}(\mathbf{R}\mathbf{R}^\mathrm{T})$ takes the minimum, there must be
\begin{equation} 
	\frac{\partial{\mathrm{tr}(\mathbf{R} \mathbf{R}^\mathrm{T})}}{\partial{\boldsymbol{u}}} = 2\mathbf{H}\boldsymbol{u} - 2\boldsymbol{f} = \mathbf{0}.
\end{equation}
Hence
\begin{equation} 
	\hat{\boldsymbol{u}} = \mathbf{H}^{-1}\boldsymbol{f}.
\end{equation}

\subsubsection{Calculation of $\mathbf{H}$ and $\boldsymbol{f}$}
It is easy to get
\begin{eqnarray} 
\mathrm{tr}(\mathbf{B}_{i_{t'},j_{t'}} \mathbf{B}_{i_{s'},j_{s'}}^\mathrm{T}) 
	&=& \mathrm{tr}(\boldsymbol{a}_{i_{t'}} \boldsymbol{a}_{j_{t'}}^\mathrm{T} \boldsymbol{a}_{j_{s'}} \boldsymbol{a}_{i_{s'}}^\mathrm{T}) \nonumber \\
	&=& \left\langle\boldsymbol{a}_{i_{t'}}, \boldsymbol{a}_{i_{s'}}\right\rangle \left\langle\boldsymbol{a}_{j_{t'}}, \boldsymbol{a}_{j_{s'}}\right\rangle.
\label{eq:H}
\end{eqnarray} 
Let $\left\langle \mathbf{B}_{i_{t'},j_{t'}}, \mathbf{B}_{i_{s'},j_{s'}} \right\rangle \triangleq 
	\left\langle \boldsymbol{a}_{i_{t'}}, \boldsymbol{a}_{i_{s'}} \right\rangle \left\langle \boldsymbol{a}_{j_{t'}}, \boldsymbol{a}_{j_{s'}} \right\rangle$,
then
\begin{equation}
	\mathbf{H} = \left(
		\begin{array}{ccc}
			\left\langle\mathbf{B}_{i_1,j_1},\mathbf{B}_{i_1,j_1}\right\rangle & \cdots & \left\langle\mathbf{B}_{i_1,j_1},\mathbf{B}_{i_t,j_t}\right\rangle\\
			\vdots 															   & \vdots & \vdots\\
			\left\langle\mathbf{B}_{i_t,j_t},\mathbf{B}_{i_1,j_1}\right\rangle & \cdots & \left\langle\mathbf{B}_{i_t,j_t},\mathbf{B}_{i_t,j_t}\right\rangle\\
		\end{array}
	\right).
\end{equation}
Similarly, for $\boldsymbol{f}$, because
\begin{equation}
	\mathrm{tr}(\mathbf{Y} \mathbf{B}_{i,j}^\mathrm{T}) = \mathrm{tr}(\mathbf{Y} \boldsymbol{a}_j \boldsymbol{a}_i^\mathrm{T}) = \boldsymbol{a}_i^\mathrm{T} \mathbf{Y} \boldsymbol{a}_j = \left\langle\mathbf{Y}, \mathbf{B}_{i,j}\right\rangle,
\end{equation}
we have
\begin{equation}
	\boldsymbol{f} = (\left\langle\mathbf{Y}, \mathbf{B}_{i_1,j_1}\right\rangle, \cdots, \left\langle\mathbf{Y}, \mathbf{B}_{i_t,j_t}\right\rangle)^\mathrm{T}.
\end{equation}

\subsection{Equivalence of 2D-OMP to 1D-OMP}
	Obviously, the $(n(i-1)+j)$-th atom of $\mathbf{\Omega}$ is
\begin{equation}
\boldsymbol{\omega}_{n(i-1)+j} = \left(
	\begin{array}{c}
		a_{1i}\boldsymbol{a}_j\\
		\vdots\\
		a_{mi}\boldsymbol{a}_j
	\end{array}
\right).
\end{equation}
Compared with (\ref{eq:2D_atom}), it can be found that $\boldsymbol{\omega}_{n(i-1)+j}$ is just the 1D stretched vector of $\mathbf{B}_{i,j}$. Hence, the Frobenius norm of $\mathbf{B}_{i,j}$ will equal the $\ell_2$-norm of $\boldsymbol{\omega}_{n(i-1)+j}$.

Then we prove that the projection of $\mathbf{R}$ onto $\mathbf{B}_{i,j}$ equals the projection of $\boldsymbol{r}$ onto $\boldsymbol{\omega}_{n(i-1)+j}$. Obviously, 
\begin{eqnarray}
	\left\langle \boldsymbol{r}, \boldsymbol{\omega}_{n(i-1)+j} \right\rangle &=& 
	\sum_{i'=1}^{m}{\sum_{j'=1}^{m}{a_{i'i}a_{j'j}r_{m(i'-1)+j'}}} \nonumber\\
	&=& \boldsymbol{a}_i^\mathrm{T} \mathbf{R} \boldsymbol{a}_j = \left\langle\mathbf{R}, \mathbf{B}_{i,j}\right\rangle.
\end{eqnarray}
Hence
\begin{equation}
	\frac{\left|\left\langle \boldsymbol{r}, \boldsymbol{\omega}_{n(i-1)+j} \right\rangle\right|}{\left\|\boldsymbol{\omega}_{n(i-1)+j}\right\|_2} = 
	\frac{\left|\left\langle \mathbf{R}, \mathbf{B}_{i,j} \right\rangle\right|}{\left\|\mathbf{B}_{i,j}\right\|_2}.
\end{equation}
It means: at each iteration of 1D-OMP and 2D-OMP, the same atom will be selected.

Finally, as $\boldsymbol{r}$ is the 1D stretched vector of $\mathbf{R}$, $\left\| \mathbf{R} \right\|_2 = \left\| \boldsymbol{r} \right\|_2$. Hence, the least squares in 1D-OMP and 2D-OMP will output exactly the same results (in fact, it can be easily proved that $\mathbf{Q} = \mathbf{H}$ and $\boldsymbol{f} = \boldsymbol{g}$).

Based on the above analysis, we draw the conclusion that 2D-OMP is equivalent to 1D-OMP.
	
\subsection{Complexity Analysis}
Below we analyze the complexity of 2D-OMP step by step.
	
\subsubsection{Project} Let $\mathbf{P}$ be an $n \times n$ matrix whose $(i,j)$-th element is $\left\|\mathbf{B}_{i,j}\right\|_2$. Then this step can be implemented by $\mathbf{A}^\mathrm{T}\mathbf{R}\mathbf{A} ./ \mathbf{P}$. The complexity of this step is dominated by $(n \times m) \times (m \times m)$ matrix-matrix multiplication and $(n \times m) \times (m \times n)$ matrix-matrix multiplication. As $m < n$, the complexity of this step is $O(mn^2)$. 

\subsubsection{Select Best Matched Unselected Atom} Since there are $n^2$ atoms and $k \ll n^2$, the complexity of this step is approximately $O(n^2)$, negligible compared with the $\textit{Project}$ step.

\subsubsection{Renew Weights} From (\ref{eq:H}), it can be seen that the complexity to calculate $\left\langle \mathbf{B}_{i_{t'}, j_{t'}}, \mathbf{B}_{i_{s'}, j_{s'}} \right\rangle$ is $O(m)$. Since $\left\langle\mathbf{Y}, \mathbf{B}_{i,j}\right\rangle = \boldsymbol{a}_i^\mathrm{T} \mathbf{Y} \boldsymbol{a}_j$, the complexity to calculate $\left\langle \mathbf{Y}, \mathbf{B}_{i,j} \right\rangle$ is $O(m^2)$. The complexity to calculate $\mathbf{H}$ and $\boldsymbol{f}$ depends on $t$. For $t \leq k \ll n^2$, the complexity of this step is negligible compared with the $\textit{Project}$ step. 

\subsubsection{Update Residual} The complexity of this step depends on $t$. For $t \leq k \ll n^2$, the complexity of this step is negligible compared with the $\textit{Project}$ step.

Based on the above analysis, we draw the conclusion that the complexity of 2D-OMP is $O(mn^2)$, roughly $1/m$ of that of 1D-OMP.

\subsection{Memory Usage Analysis}
For 1D-OMP, an $m^2 \times n^2$ matrix is needed to hold $\mathbf{\Omega}$, so the memory usage is $O(m^2n^2)$, while for 2D-OMP, $\mathbf{\Omega}$ is replaced by $\mathbf{A}$, so the memory usage is reduced to $O(mn)$.

\section{Experimental Results}\label{sec:results}
\begin{figure}
\centering
\includegraphics[width=.5\linewidth]{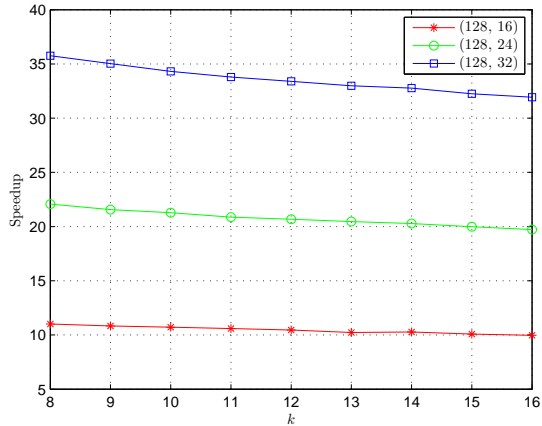}
\caption{Speedup of 2D-OMP over 1D-OMP.}
\label{fig:speedup}
\end{figure}

	We have written both 2D-OMP and 1D-OMP algorithms in MATLAB \cite{Homepage}. We present herein the results under three typical settings, i.e. $(n, m) = (128, 16)$, $(128, 24)$, and $(128, 32)$. For each setting, we increase sparsity level $k$ from 8 to 16. The transform matrix $\mathbf{\Psi}$ is $128 \times 128$ 2D discrete cosine transform (DCT) matrix. The sensing matrix $\mathbf{\Phi}$ is formed by sampling independent and identically-distributed (i.i.d.) entries from standard Gaussian distribution by using the $randn$ function. 2D Sparse signal is obtained by using the $sprandn$ function with density $k/n^2$.
	
	We run the MATLAB codes on Intel(R) Core(TM) i7 CPU with 12GB memory and collect the total running time of $10^3$ trials for 1D-OMP and 2D-OMP respectively. Because two algorithms output exactly the same results, only the speedup of 2D-OMP over 1D-OMP with respect to $k$ is reported in Fig. \ref{fig:speedup}. From Fig. \ref{fig:speedup}, we can draw two conclusions:
\begin{enumerate}
	\item As $k$ increases, the speedup of 2D-OMP over 1D-OMP descends gradually. This is because for small $k$, the complexity of 2D-OMP and 1D-OMP is dominated by the \textit{Project} step. As $k$ increases, the complexity of other steps will weight heavier. Especially, at the \textit{Renew Weights} step, the complexity of $t \times t$  matrix inverse is $O(t^3)$, which will ascend quickly as $t$ increases.  
	\item As $m$ increases, the speedup of 2D-OMP over 1D-OMP becomes more significant. When $m = 16$, the speedup of 2D-OMP over 1D-OMP ranges from 10 to 11 times, while when $m = 32$, the speedup of 2D-OMP over 1D-OMP ranges from 32 to 35 times. This is because the speedup of 2D-OMP over 1D-OMP comes mainly from the \textit{Project} step, while at other steps 2D-OMP shows little superiority over the 1D-OMP. As $m$ increases, the complexity of the \textit{Project} step weights heavier, which explains the aove phenomenon. 
\end{enumerate}

%%%%%%%%%%%%%%%%%%%%%%%%%%%%%%%%%%%%%%%%%%%%%%%%%%%%%%%%%%%%%%%%%%%%%%%%%%%%%%%%%%%%%%%%%%%%%%%%%%%%%%%%%%%%%%%%%%%%%%%%%%%%%%%%%%%%%%%%%%%%%%%%%%%%%%%%%%%%%%%%%%%%%%%%%
\section{Conclusion}\label{sec:conclusion}
	For 2D sparse signal recovery, this paper develops 2D-OMP algorithm. We prove that 2D-OMP is equivalent to 1D-OMP, but it reduces recovery complexity and memory usage. Hence, 2D-OMP can be used as an alternative to 1D-OMP in such scenarios as compressive imaging, image compression, etc.
	
	Following the deduction in this paper, the extension of 2D-OMP to higher dimensional OMP is straightforward. For example, by utilizing 3D separable sampling, 3D-OMP can be obtained by defining each atom as a 3D matrix. Then at each iteration, the decoder projects 3D sample matrix onto 3D atoms to select the best matched atom, and then renews the weights for all the already selected atoms via the least squares. 3D-OMP can find its use in hyperspectral image compression.

%%%%%%%%%%%%%%%%%%%%%%%%%%%%%%%%%%%%%%%%%%%%%%%%%%%%%%%%%%%%%%%%%%%%%%%%%%%%%%%%%%%%%%%%%%%%%%%%%%%%%%%%%%%%%%%%%%%%%%%%%%%%%%%%%%%%%%%%%%%%%%%%%%%%%%%%%%%%%%%%%%%%%%%%%

%\begin{biography}[{\includegraphics[width=1in,height=1.25in,clip,keepaspectratio]{eps/fangyong.eps}}]
%{Yong Fang} received his BEng, MEng, and PhD from Xidian University, Xi'an China, in 2000, 2003, and 2005, respectively, all in signal processing. Then, he was appointed as a 
%post-doctoral fellow for one year at Northwest Polytechnique University, Xi'an China. From 2007 to 2008, he joined Hanyang University, Seoul Korea, as a research %professor. He is currently a professor at Northwest A\&F University, Shaanxi Yangling, China. He has long experiences in hardware development, e.g. FPGA-based (Xilinx %Vertex series) video codec design, DSP-based (TI C64 series) video surveilliance system, etc. His research interests include distributed source coding, image/video %coding, processing, and transmission.
%\end{biography}

\end{document}